\documentclass[conference,onecolumn]{IEEEtran}

\usepackage{geometry}
\newtheorem{thm}{Theorem}
\newtheorem{remk}{Remark}

\usepackage[final]{graphicx}
\usepackage[reqno]{amsmath}
\usepackage{amssymb}
\usepackage{subfig}
\usepackage{epstopdf}
\usepackage{bm}
\usepackage{subfig}
\begin{document}

\sloppy
\title{DoF Analysis in a Two-Layered Heterogeneous Wireless Interference Network}

\author{
    \IEEEauthorblockN{Meghana Bande, Venugopal V.~Veeravalli }
    \IEEEauthorblockA{ECE Department and CSL\\ University of Illinois at Urbana-Champaign \\ Email: \{mbande2,vvv\}@illinois.edu}
    \and
     \IEEEauthorblockN{Antti T{\"o}lli, Markku Juntti}
    \IEEEauthorblockA{Centre for Wireless Communication,\\ University of Oulu, Finland\\
Email: \{antti.tolli, markku.juntti\}@ee.oulu.fi\\
}}
\maketitle

\begin{abstract}
Degrees of freedom (DoF) is studied in the downlink of a heterogenous wireless network modeled as a two-layered interference network. The first layer of the interference network is the backhaul layer between macro base stations (MB) and small cell base stations (SB), which is modeled as a Wyner type linear network. The second layer is the transmission layer between SBs and mobile terminals (MTs), which is modeled as a linear Wyner $L_T$ network, i.e., each MT is connected to $L_T+1$ SBs. The SBs are assumed to be half-duplex, thus restricting the per user degrees of freedom (puDoF) in the system to $1/2$. The puDoF can be further restricted by the number of antennas at the MB.
For $L_T \in \{1,2\}$, the optimal puDoF can be achieved by using simple interference avoidance schemes.
The increase in the connectivity of transmission layer beyond $L_T=2$ limits the achievable puDoF using only zero-forcing schemes to less than 1/2, even in the presence of large number of antennas at each MB but the optimal puDoF can be achieved by making each message available at multiple SBs. 
This is done by sending an appropriate linear combination to the SB to zero-force interference at the intended user. 
The maximum per user DoF of 1/2 can be achieved in the linear network with sufficient number of antennas using only interference avoidance schemes. These results are also extended to a more realistic hexagonal cellular model as well. 
 \end{abstract}

\section{Introduction}

To meet the increasing demand for mobile traffic, heterogeneous networks are envisioned to be a key component
of future cellular networks \cite{Hetnet1}. Heterogeneous networks enable flexible and low-cost deployments and provide a uniform broadband experience to users anywhere in the network \cite{Hetnet2}. Managing interference in heterogeneous networks is crucial in order to achieve higher data rates for users. We consider the downlink of a heterogeneous network with $K_B$ macro base stations (MBs), $K$ small cell base stations (SBs) and $K$ mobile terminals (MTs) with $K = S\times K_B$ for an appropriate $S$. It is assumed that the MBs and the SBs operate on the same frequency band. 

The SBs act as relays between MBs and MTs. We consider a linear interference model for both the backhaul and the transmission layer. The channel between the macro and the SBs is modeled as a Wyner type network while the channel between SBs and MTs is modeled as a Wyner $L_T$ model. The dependence of DoF in this network on several factors such as the number of antennas at the MB, interference in the backhaul and transmission layers is investigated in this work. These insights are applied to a practical model with the transmission layer modeled as a hexagonal cellular network. 

The DoF in locally connected linear networks was studied in~\cite{Lapidoth-Shamai-Wigger-ISIT07},~\cite{Shamai-Wigger-ISIT11},~\cite{ElGamal-Annapureddy-Veeravalli'14} using cooperation under maximum transmit set size cooperation constraints. We use insights from these works to characterize the DoF in the transmission layer. 
In the schemes of \cite{Lapidoth-Shamai-Wigger-ISIT07}-\cite{gamal2013dynamic}, the messages of multiple users are available at some of the transmitters. This requires multiple time-slots in the backhaul layer in our setting. The key observation here is that the SBs use the knowledge of multiple messages to null the interference at the intended MT. Hence it is sufficient if the MBs send linear combinations of the messages to the SBs. This would require that at each MB, the channel between SBs and the corresponding MTs is known. 

A two-layered interference network modeled as a $K\times K\times K$ relay channel with each layer as a $K$-user interference channel with full connectivity was considered in \cite{Salman}. Using aligned-network-diagonalization, the maximum sum-DoF of $K$ was achieved. The sum-DoF was studied for the case of $K=2$  i.e., $2 \times 2 \times 2$ in \cite{SalmanLinear} under restriction to linear schemes, and was shown to be $4/3$. In contrast to these schemes, we consider a broadcast channel in the first layer and local connectivity in both layers. We also restrict ourselves to more practical zero-forcing schemes. To the best of our knowledge, there has been no prior work on a DoF analysis for a two-layered network of the kind we consider here.

The system model is described in Section \ref{sec:systemmodel}.
 The DoF analysis for $L_T\in\{1,2\}$ is presented in Section \ref{sec:nocoop}. 
 The DoF analysis for the case of a general $L_T $, for which a per user DoF (puDoF) of 1/2 can be achieved with zero-forcing schemes only by sending linear combinations from MB's to corresponding SB's is presented in Section \ref{sec:gen}. The system model for the cellular network and the results are discussed in Section \ref{sec:cellular} and finally the concluding remarks are presented in Section \ref{sec:conc}.
 
\section{System Model and Notation}\label{sec:systemmodel}

We consider a heterogeneous wireless network with MBs, SBs and MTs. It is assumed that MBs do not directly serve MTs and do not cause interference at MTs. We assume that the SBs act as half-duplex relays between the MBs and the MTs.
There are two layers in this network, the \emph{backhaul layer} between MBs and SBs and the \emph{transmission layer} between SBs and MTs.

\subsection{Transmission Layer}

Consider the transmission layer with $K$ SBs and MTs. Let ${\cal K}$ denote the set $\{1,...,K\}$. Each SB is equipped with a single antenna. In the transmission layer, the channel gain between SB $j, \forall j \in {\cal K}$ and MT $i, \forall i \in {\cal K}$ is denoted by $h^{\text{Tx}}_{ji}$.
At each MT $i$, the received signal $y^{\text{Tx}}_i$ is given by
\begin{equation}\label{signal}
y^{\text{Tx}}_{i}(t)=h^{\text{Tx}}_{ii}(t)x_{i}^{\text{Tx}}(t)+\sum_{j\in {\mathcal I}_{i}}h^{\text{Tx}}_{ji}(t)x^{\text{Tx}}_{j}(t)+z^{\text{Tx}}_{i}(t),
\end{equation}
where $t$ denotes the time-slot, $x^{\text{Tx}}_{j}(t)$ denotes the signal transmitted by SB $j$
under an average transmit power constraint, $z_{i}^{\text{Tx}}(t)$ denotes the additive white
Gaussian noise at MT $i$, $h^{\text{Tx}}_{ji}(t)$ denotes the channel
gain coefficient from SB $j$ to MT $i$, and ${\mathcal I}_{i}$ denotes
the set of interferers at MT $i$.  

The cellular model presented by Wyner \cite{Wyner} was extended in \cite{ElGamal-Annapureddy-Veeravalli'14} to a locally connected linear interference network with connectivity parameter $L_T$. The transmission layer is assumed to be a local Wyner $L_T$ model with $K$ users. The cells are located on an infinite linear equi-spaced grid and each transmitter is associated with a single user. Here $L_T$ denotes the number of dominant interferers per user, where each user observes interference from $\lceil \frac{L_T}{2}\rceil$ preceding and $\lfloor \frac{L_T}{2}\rfloor$ succeeding transmitters. 
The Wyner $L_T$ model is illustrated in Figure \ref{fig_wynach} and the channel coefficients are given by
\begin{equation*}
 h^{\text{Tx}}_{ji}(t)\neq 0  \text{ iff } i\in\{j - \lfloor \frac{L_T}{2}\rfloor, \ldots, j - 1, j, j + 1, \ldots,j + \lceil \frac{L_T}{2}\rceil\}.
\end{equation*}

 \begin{figure}[htb]
\centering
\includegraphics[scale=0.45]{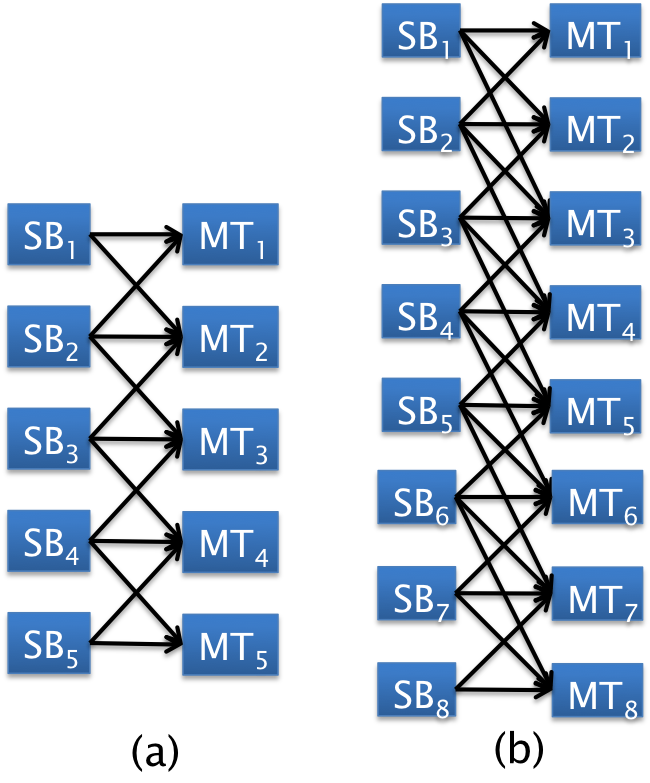}
\protect\caption{Wyner interference network in the transmission layer with (a) $L_T=2$ (b) $L_T=3$.}
\label{fig_wyner_sysmodel}
\end{figure}

\subsection{Backhaul Layer}

The backhaul layer is assumed to be a linear model with connectivity $L_B$. Let ${\mathcal S}_i$ denote the set of $S$ consecutive SBs where ${\mathcal S}_i=\{(i-1)S+1, \ldots, (i)S\}$. Each MB $i$ is associated with a set ${\mathcal A}_i$ of $S+L_B$ consecutive SBs where ${\mathcal A}_i = {\mathcal S}_{i-1}(\{S-\lfloor \frac{L_B}{2}\rfloor,\ldots,S\})\cup {\mathcal S}_i \cup {\mathcal S}_{i+1}(\{1,\ldots, \lceil \frac{L_B}{2}\rceil\ \})$. Transmission from MB $i$ to any SB in ${\mathcal S}_i$ causes interference at $\lfloor \frac{L_B}{2}\rfloor$ SBs above and at $\lceil \frac{L_B}{2}\rceil$ SBs below the set ${\mathcal S}_i$.  Let $N$ denote the number of antennas at each MB. 

The channel vector between MB $i$ and SB $j$ at time-slot $t$ is denoted by $ \mathbf{h}^{B}_{i,j}(t)$. The channel model for backhaul layer is given by $ \mathbf{h}^{B}_{i,j}(t)\neq 0$ iff  $j\in {\mathcal A}_i$. Let the channel gain matrix corresponding to MB $i$, $\mathbf{H}^{B}_{i}(t) \in \mathbb{C}^{(N) \times (S+L_B)} = [{\mathbf h}^{B}_{i, \mathcal{S}_{i-1}(S-\lfloor \frac{L_B}{2}\rfloor)}(t), ..., {\mathbf h}^{B}_{i, \mathcal{S}_i(S)}(t),\ldots, {\mathbf h}^{B}_{i, \mathcal{S}_{i+1}(\lceil \frac{L_B}{2}\rceil)}(t)]$ in the backhaul layer where the $j$th column corresponds to the channel coefficients from MB $i$ to SB $j$. Also let ${\mathbf x}_{i}^{B}(t) \in \mathbb{C}^{N\times 1}$ to be the transmitted signal vector from MB $i$ and $z_{k}^{B}(t)$ denotes the additive white
Gaussian noise at SB $k$. The received signal at $k$th SB served by MB $i$ is given by,
\[
y^{B}_k(t)= ({\mathbf h}^{B}_{i,k}(t))^{T}{\mathbf x}^{B}_{i}(t) + \sum_{j\neq i}  ({\mathbf h}^{B}_{j,k}(t))^{T}{\mathbf x}^{B}_{j}(t) + z^{B}_k(t).
\]
 Let ${\cal R}_{i}(t)\subseteq {\mathcal A}_i$ denote the set of SBs receiving messages from MB $i$ in a particular time-slot $t$.
 
\begin{figure}[htb]
\centering
\includegraphics[scale=0.4]{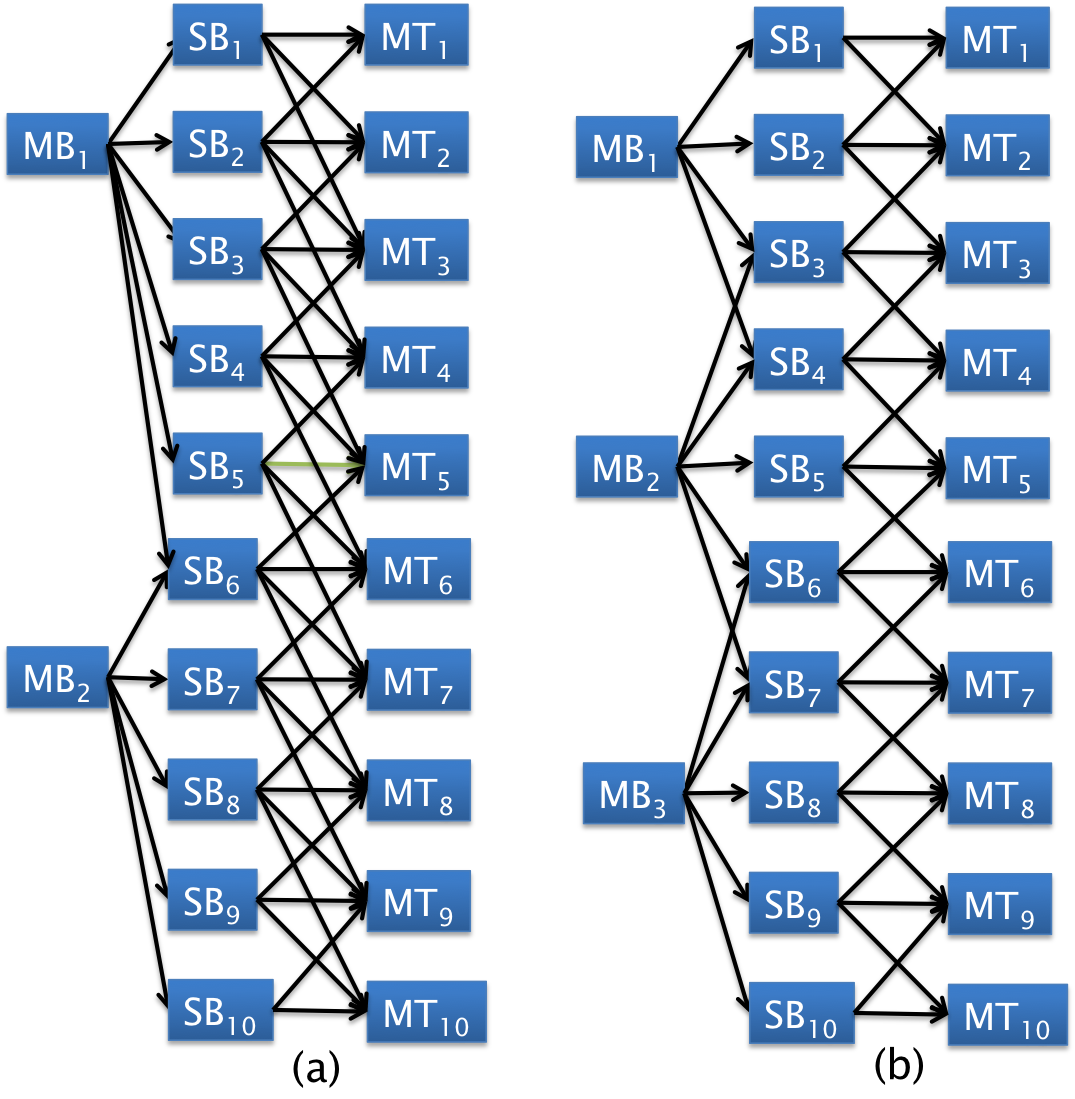}
\protect\caption{Two-layered network with (a) $S=5$ and $L_B=1$ in the backhaul layer and $L_T=3$ in the transmission layer and (b) $S=3$ and $L_B=2$ in the backhaul layer and $L_T=2$ in the transmission layer.}
\label{fig_total_sysmod}
\end{figure}

Local channel state information is assumed to be available at MBs and SBs. All channel coefficients that are not identically zero are assumed to be drawn independently from a continuous distribution.
The system model is illustrated in Figure \ref{fig_total_sysmod}. 
 
\subsection{Capacity and Degrees of Freedom}
Let $P$ be the average transmit power constraint at each SB and the transmit power per antenna at a MB. Let ${\cal W}_{i}$ denote the alphabet for $W_{i}$, where $W_i$ denotes the message for MT $i$. The rates
$R_{i}(P)=\frac{\text{log}|{\cal W}_{i}|}{n}$ are achievable iff the error probabilities of all messages can simultaneously be arbitrarily small for large $n$, using an interference management scheme.
The degree
of freedom (DoF) $d_{i}, \forall i \in {\cal K}$ is defined as 
\begin{equation}
d_{i}=\underset{P\rightarrow\infty}{\text{lim}}\frac{R_{i}(P)}{\text{log}P}.
\end{equation}
 DoF is  the number of interference free sessions in a multi-user channel at asymptotically high signal-to-noise ratio (SNR).
 The maximum achievable sum DoF $\eta(K)$ in a channel with $K$ users (MTs) 
 is defined as
$\eta (K) = \underset{{\cal D}}{\text{max}}\underset{i\in{\cal K}}{\text{\ensuremath{\sum}}}d_{i}$. 
where ${\cal }$${\cal D}$
denotes the closure of the set of all achievable DoF tuples.
 Then the maximum achievable per user DoF (puDoF) is defined as
\begin{equation}
\tau=\underset{K\rightarrow\infty}{\text{lim}}\frac{\eta(K)}{K}.
\end{equation}

\section{DoF Analysis for $L_T=\{1,2\}$}\label{sec:nocoop}

In this section, we consider the case where the connectivity in the transmission layer $L_T\leq 2$ while $L_B=1$. We present upper bounds on the achievable puDoF which hold for general $L_T$ and $L_B$. Similar achievability schemes can be used for higher values of $L_T$ and $L_B$.

Note that at any MB $i$, $N_1+1$ antennas are sufficient in order to send messages to $N_1$ SBs ($|{\cal R}_{i}|=N_1$) and to null the interference at SB $z$. Let $\mathbf{X}\in\mathbb{C}^{N_1+1}$ denote the transmitted signal vector at MB $i$, $\mathbf{H}$ denote $[\mathbf{H}^{B}_{i, {\cal R}_i}, \mathbf{h}^{B}_{i, z}]$, and $\mathbf{W} \in \mathbb{C}^{N_1+1}$ denote the vector containing the intended messages to ${\cal R}_{i}$ appended with zero at the end. Then we have $\mathbf{HX}^{T} = \mathbf{W}^{T}$. From our assumptions, $\mathbf{H}$ is full rank almost surely and a solution for $\mathbf{X}$ is obtained.

We present the following theorem for the case $L_T\in\{1,2\}$.
\begin{thm} \label{thm_nocoop}
The following bounds hold for $\tau$, when $L_B=1$, $L_T\in\{1,2\}$, 
\begin{align} \label{eqn_ach1}
\tau \geq \left\{ \begin{array} {cl}
\frac{N}{S} & \mbox{for $N < S/2$} \\
\frac{1}{2}(1-\frac{1}{S}) & \mbox{for $N = S/2$ for $S$ even} \\
\frac{1}{2} & \mbox{for $N > S/2$}
\end{array} \right. 
\end{align}

\begin{equation}\label{eqn_con1}
\tau \leq \textup{min}(\frac{N}{S},\frac{1}{2}).
\end{equation}

\end{thm} 

\begin{IEEEproof}
We present the achievable scheme in order to show the bounds  (\ref{eqn_ach1}). In the transmission layer for Wyner model with $L_T\in \{1,2\}$, by deactivating alternate transceiver pairs, the remaining messages can be sent interference free as shown in Figure \ref{fig_wynach}. Thus, a puDoF of 1/2 is achieved if the corresponding messages are available at the active SBs. 

Case 1: $N > \frac{S}{2}$.

A) When $S$ is odd, our achievable scheme uses only $\frac{S+1}{2}$ antennas at an MB. Consider the following message assignment for each time-slot $t$ where $t$ is odd.   
    \[
    {\cal R}_i(t) = \begin{array}{ll}
        {\mathcal S}_i(\{1, 3, ..., S\}) & \textup{for }i \textup{ odd}\\   
        {\mathcal S}_i(\{2, 4, ..., S-1\}) & \textup{for }i \textup{ even.}                          
        \end{array}
  \]    
\begin{figure}[htb]
\centering
\includegraphics[scale=0.4]{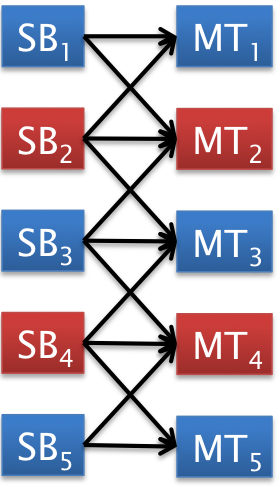}
\protect\caption{Achievable schemes in the transmission layer with $L_T=2$ and puDoF $=1/2$. The red boxes indicate deactivated transceivers.}
\label{fig_wynach}
\end{figure}  

%(b)  $L_T=3$ and  puDoF $=1/2$.

When $i$ is even, SB ${\mathcal S}_i(1)$ is not active in this time-slot. Only when $i$ is odd, ${\mathcal S}_i(1)$ observes interference from the transmissions of MB $i-1$. MB $i-1$ needs $ \frac{S-1}{2}$ antennas for sending messages and one antenna for nulling the interference at SB ${{\mathcal S}_i(1)}$. Thus at the end of each odd time-slot, messages are available at alternate SBs and a puDoF of $1/2$ is achieved.
The assignment is reversed when $t$ is even and the achievability follows similarly.

B) When $S$ is even, our achievable scheme uses $\frac{S}{2}+1$ antennas at an MB. In odd and even time-slots, only the odd and even numbered SBs are served respectively. This is possible as the cluster of any MB $i$ contains $\frac{S}{2}$ SBs with odd and $\frac{S}{2}$ SBs with even indices.
Only in time-slots $t$, where $t$ is odd the first SB in each active cluster observes interference. Each MB uses $\frac{S}{2}$  antennas to send messages and has an additional antenna to null interference at the first SB in the next cluster. Thus, in each time-slot, messages are available at alternate SBs and a puDoF of $1/2$ is achieved.

Case 2: $N < \frac{S}{2}$. In this case, $S \geq 2N+2$ or $S \geq 2N+1 $. Hence in each cluster, two disjoint sets of $N$ SBs are served in consecutive time-slots while the first SB of the cluster is inactive.
Consider the following message assignment for each time-slot $t$ when $t$ is odd.
  \[
    {\cal R}_i(t) = \begin{array}{ll}
        {\mathcal S}_i(\{3, 5, ..., 2N+1\}) & \textup{for }i \textup{ odd}\\   
        {\mathcal S}_i(\{2, 4, ..., 2N\}) & \textup{for }i \textup{ even.}                          
        \end{array}
  \]   
  This assignment is reversed when $t$ is even. 
The first SB in each cluster is not served at all and hence there is no interference in the backhaul layer. In each time-slot, $N$ messages among every $S$ users are sent interference-free, achieving a puDoF of $\frac{N}{S}$.

Case 3: $N = \frac{S}{2}$.
This case arises only when $S$ is even. For an even time-slot $t$, let the even numbered SBs be served. Only the first SB in each cluster sees interference, and hence there is no interference in the backhaul layer in this time-slot. When $t$ is odd, all the odd numbered indices ($S/2-1$) except for the first ones in each cluster are served. In the transmission layer, these messages are sent interference free and a puDoF of $\frac{1}{2}(\frac{1}{2}+\frac{S/2-1}{S})$ is achieved.

For the converse, we prove a stronger result that holds for any $L_B,L_T$ and any number of antennas at the SBs.
\begin{equation}\label{eqn_con}
 \tau \leq \textup{min} (\frac{N}{S},\frac{1}{2}).
\end{equation}

Consider any SB $i$. For every message SB $i$ sends, a time-slot was required in receiving that message in the backhaul layer due to the half duplex nature. 
After $T$ time-slots, the maximum number of messages transmitted by each SB $i$ is $T/2$ and received by all MTs is $K(T-1)/2$.
 For any scheme,
\[
 \textup{puDoF}=\frac{\textup{No. of messages received interference-free}}{KT} \leq \frac{1}{2}.
\] 

The number of macro base stations is $\lceil \frac{K}{S} \rceil$. 
After $T$ time-slots, the maximum number of messages that can be received by the SBs is $\lceil \frac{K}{S} \rceil T N$ messages. Hence the maximum number of messages that can be received by the mobile terminals is $\lceil \frac{K}{S} \rceil (T-1) N$.
For any scheme,
\[
 \textup{puDoF}\leq (\frac{1}{S}+\frac{1}{K})N(1-\frac{1}{T}).
 \]
 This approaches $ \frac{N}{S}$ when $T$ and $K$ become large.
\end{IEEEproof}
 
\begin{figure}[htb]
\centering
\includegraphics[scale=0.4]{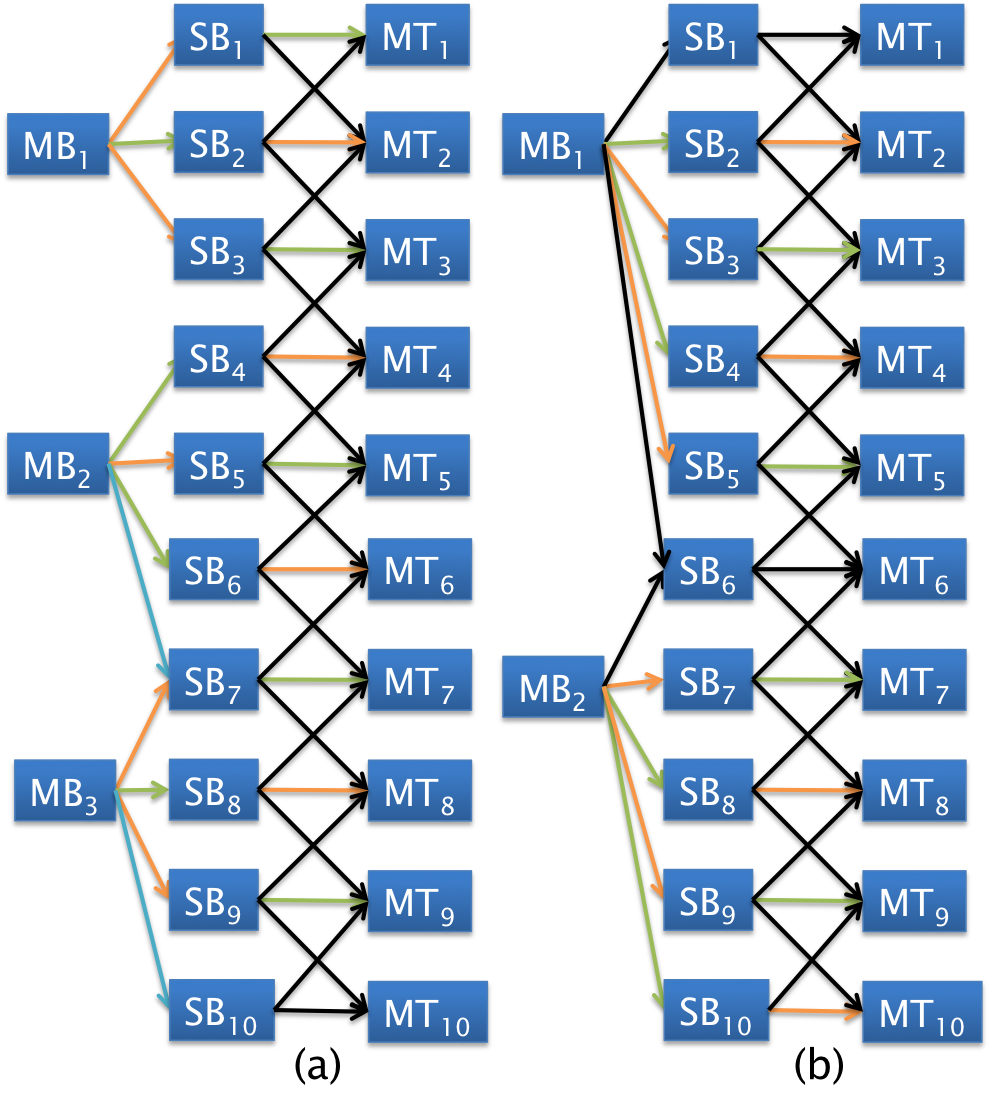}
\protect\caption{Achievable schemes for the network with $L_B=1$ and $L_T=2$ (a) puDoF $=1/2$ with $S=3$ and $N=2$ (b) puDoF $=2/5$ with $S=5$ and $N=2$. The green and orange arrows indicate the transmissions in consecutive time-slots and blue, the nulling beam.}
\label{fig_ach1}
\end{figure}

The upper bounds in Equation (\ref{eqn_con1}) hold in general for all possible achievable schemes, but the maximum puDoF can be achieved by simple interference avoidance schemes except for the case $N = \frac{S}{2}$ when $L_T \in \{1,2\}$. The achievable schemes are illustrated in Figure \ref{fig_ach1}. Even for a general $L_B$, by employing a sufficiently large number of antennas $N \geq \lceil\frac{S}{2}\rceil+L_{B}$ at the macro base stations, the interference in backhaul layer can be eliminated and a puDoF of $1/2$ can be achieved.

\section{Achievable schemes for general $L_T$}\label{sec:gen}

 The optimal puDoF for a given number of antennas cannot be achieved for higher values of $L_T$ using only interference avoidance schemes without the use of cooperation. For example, when $L_T=3$, with restriction to only ZF schemes without cooperation at the SBs, we have $\tau \leq \frac{2}{5}$ in the transmission layer even for a large $N$ from \cite{ElGamal-Annapureddy-Veeravalli'14}. 
 We consider cooperation among the SBs and show that the optimal puDoF can be achieved for $L_T\in \{3,4\}$ using only interference avoidance schemes. For cooperation, multiple messages need to be available at SBs for transmission in a particular time-slot. This requires multiple time-slots for transmission by the MBs in the backhaul layer which leads to ineffective use of resources. The SBs use the knowledge of messages available only for zero-forcing, and, thus, it suffices to have a linear combination of messages at the SBs. This would require only one time-slot for transmission in the backhaul layer. However, this would require that at each MB, the channel between SBÕs and the corresponding MTs is known. While this might require a large amount of CSI to be present at each MB, this would be justified if the coherence time is large enough.

\begin{remk}\label{rem:1}
In the Wyner $L_T$ model, if groups of $A$ SB-MT pairs are separated by $F$ where $F \geq \lceil \frac{L_T}{2}\rceil$ deactivated pairs, then there is no interference between the groups. If all $A$ messages are sent such that the interference at each MT is zero-forced, a puDoF of $A/(F+A)$ is achieved if the messages are available at the SBs.
\end{remk}

\begin{thm}\label{thm_gen1}
For $L_B=1$, if $S \geq \lceil \frac{L_T}{2}\rceil$ and $N \geq S$, we have 
$
\tau \geq \frac{1}{2}.
$
\end{thm}

\begin{IEEEproof}
In every even slot $t$, all the SBs belonging to ${\cal S}_i$ receive linear combinations of messages for $i$ even and the remaining SBs receive in the odd time-slot $t$. This is possible since we have $N \geq S$ antennas.
Thus, in each time-slot, groups of $S$ SBs separated by $S$ SB-MT pairs send messages interference free to their respective MTs. A puDoF of $1/2$ is achieved.
\end{IEEEproof}

\begin{thm}\label{thm_gen2}
The following bounds hold for $\tau$, when $L_B=1$ when $\lfloor \frac{S}{2}\rfloor \geq \lceil \frac{L_T}{2}\rceil$,
\begin{align} 
\tau \geq \left\{ \begin{array} {cl}
\frac{N}{S} & \mbox{for $N < \frac{S}{2}$} \\
\frac{1}{2}(1-\frac{1}{S}) & \mbox{for $N = \frac{S}{2}$} \\
\frac{1}{2} & \mbox{otherwise.}
\end{array} \right. 
\end{align}
\end{thm}

\begin{figure}[htb]
  \centering
\subfloat[]{\label{fig:two-dim}\includegraphics[width=0.20\textwidth]{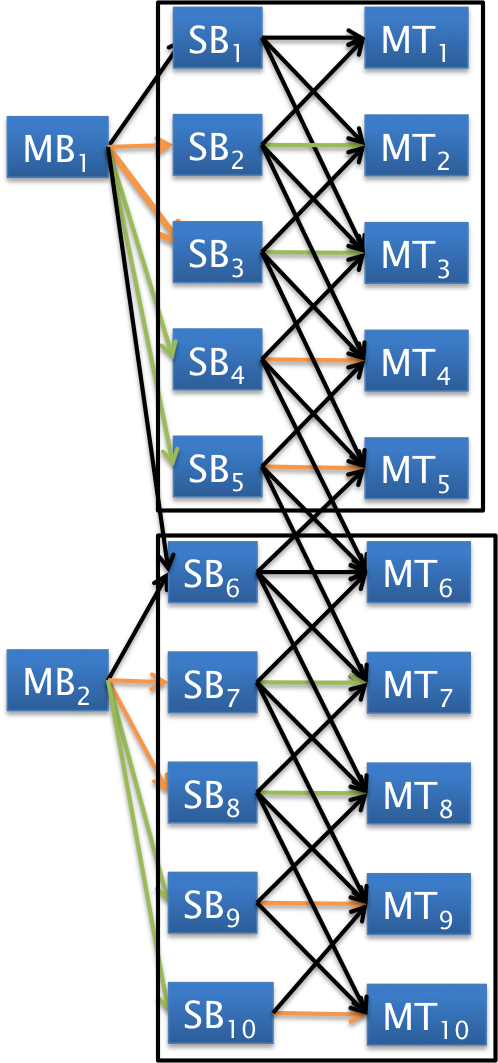}}                
\quad\quad\subfloat[]{\label{fig:twodim-codingscheme}\includegraphics[width=0.18\textwidth]{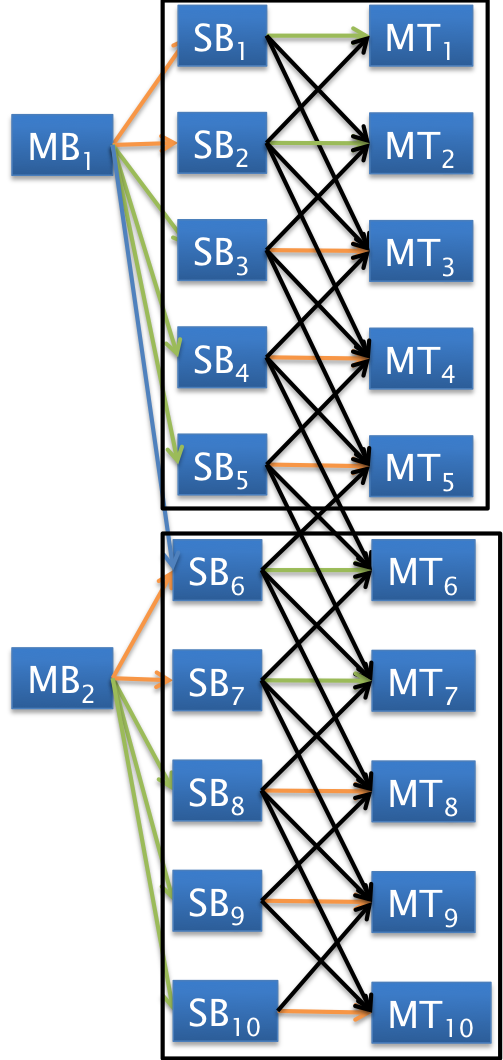}}
  \caption{Achievable schemes for the network with $L_B=1$, $L_T = 3$ and $S = 5$. In (a), $N=2$, $N<S/2$ and puDoF~$=2/5$ is achieved. In (b), $N=3$, $N>S/2$ and puDoF $=1/2$ is achieved. The green and orange arrows indicate transmissions in consecutive time-slots and blue, the nulling beam.}.
  \label{fig_ach3}
\end{figure}

\begin{IEEEproof}

A) $N > \frac{S}{2}$ is same as $N \geq \lfloor \frac{S}{2}\rfloor+1$. For all $i$, let
  \[
    {\cal R}_i(t) = \begin{array}{ll}
        {\mathcal S}_i(\{1,\ldots,\lfloor \frac{S}{2}\rfloor\}) & \textup{for }t \textup{ odd}\\   
        {\mathcal S}_i(\{\lfloor \frac{S}{2}\rfloor+1,\ldots,S\}) & \textup{for }t \textup{ even.}                          
        \end{array}
  \]  
In even and odd time-slots, $\lfloor \frac{S}{2}\rfloor+1$ and $\lfloor \frac{S}{2}\rfloor$ antennas respectively at each MB $i$ are used to send linear combinations to the SBs and in an odd time-slot one antenna is used to ZF interference at ${\cal S}_{i+1}(1)$. From the remark, the puDoF is $1/2$.

B) $N < \frac{S}{2}$ is same as $N \leq \lceil \frac{S}{2}\rceil-1$. For all $i$, let
  \[
    {\cal R}_i(t) = \begin{array}{ll}
        {\mathcal S}_i(\{2,\ldots,N+1\}) & \textup{for }t \textup{ odd}\\   
        {\mathcal S}_i(\{\lceil \frac{S}{2}\rceil+1,\ldots,\lceil \frac{S}{2}\rceil+1+N\}) & \textup{for }t \textup{ even.}                          
        \end{array}
  \]    
  In each time-slot, $N$ antennas at each MB $i$ are used to send linear combinations to the SBs. The first SB in each cluster is always inactive. Each group of $N$ SBs are separated by $S-N$ SBs and hence from the remark puDoF is $N/S$.
  
C)  $N = \frac{S}{2}$. This case arises when $S$ is even.
For all $i$, 
  \[
    {\cal R}_i(t) = \begin{array}{ll}
        {\mathcal S}_i(\{2,\ldots,\frac{S}{2}\}) & \textup{for }t \textup{ odd}\\   
        {\mathcal S}_i(\{ \frac{S}{2}+1,\ldots,S\}) & \textup{for }t \textup{ even.}                          
        \end{array}
  \]  
In the odd and even time-slots, $N-1$ and $N$ antennas respectively at each MB $i$ are used to send linear combinations to the SBs. Hence we have puDoF of $N/S$ and $N-1/S$ in consecutive time-slots, giving an average puDoF of $\frac{2N-1}{2S}$.

\end{IEEEproof}

%We have given achievable schemes for $L_T=\{1,2\}$ for $S \geq 2 $ in Theorem \ref{thm_gen2} and $S=1$ in Theorem \ref{thm_gen1}.

The achievable schemes are illustrated in Figure \ref{fig_ach3}. 
We note that for a general $L_B$, a puDoF of 1/2 can be
achieved by employing a sufficient number of antennas ($N \geq \lfloor \frac{S}{2}\rfloor+L_B $).

\section{Discussion}
\subsection{Fairness}
In our achievable schemes, we would like to discuss the notion of fairness since few of the SBs and MTs are being deactivated. We first observe that in any scheme achieving a per user DoF of half, all of the MTs receive their message once every two time-slots on average. In the case where the number of antennas is not sufficient i.e., $N<S/2$, the schemes discuss deactivating some of the users. It is easy to see that fairness can be maintained over
all users through fractional reuse in the system by deactivating different sets of receivers in different time slots. The schemes require that the first SB $i$ in each cluster to be inactive. In order to facilitate this, we use some other SB to transmit to the MT $i$ in each cluster if needed. In each time-slot, since we look at blocks separated by at least $\lceil L_T/2\rceil$, this is feasible. For the case of $L_T=\{1,2\}$, in order to we achieve fairness, we would need to send linear combinations to the appropriate SBs.
% Fairness can be maintained in the allocation of the available DoF over
%all users through fractional reuse in a system by deactivating different sets of receivers in different sessions, e.g., in different time or frequency slots.

\subsection{Massive MIMO}
We believe that with the emergence of massive MIMO, a large number of antennas at an MB is not inconceivable \cite{massive}. From the results on linear networks, we observe the theoretic upper bound of a half can be achieved by simple ZF schemes if appropriate number of antennas are made available at the MBs.

\section{Hexagonal cellular network}\label{sec:cellular}

We begin this section by first describing the transmission and backhaul layer model for the hexagonal sectored cellular network.

\subsection{Transmission Layer}

\begin{figure}[htb]
\centering
\includegraphics[scale=0.6]{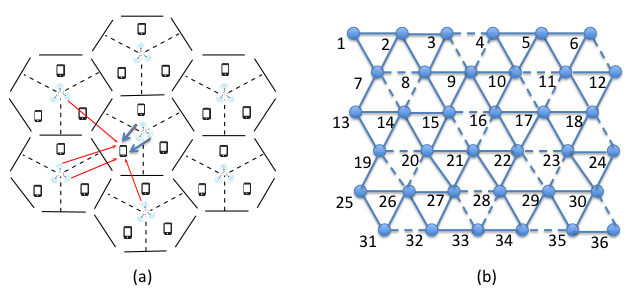}
\protect\caption{Cellular Network (a) and Interference Graph (b). The dotted lines in (b) represent interference between sectors belonging to the same cell.}
\label{Cellular Network}
\end{figure}

For the transmission layer, we consider a large sectored $K$ user cellular network with three sectors per cell as shown in Figure \ref{Cellular Network}a). A local interference model is assumed, where the interference at each receiver is only due to the base stations in the neighboring sectors. It is assumed that sectors belonging to the same cell do not interfere with each other. This is motivated by the fact that the interference power due to sectors in the same cell is usually far lower than the interference from out-of-cell users located in the sectorÕs line of sight.

\subsubsection{Interference graph}
The cellular model is represented by an undirected interference graph
$G(V,E)$ shown in Figure \ref{Cellular Network}(b) where each vertex $u\in V$ corresponds
to a transmitter-receiver pair. For any node $a$, the transmitter, receiver and intended message corresponding to the node are denoted by $T_{a}$, $R_{a}$ and $W_{a}$. An edge $e\in E$ between two vertices  $u,v\in V$ corresponds
to interference between the transmit-receiver pairs i.e., transmitter at $u$ causes
interference at the receiver corresponding to $v$ and vice-versa. The dotted lines denote interference between sectors that belong to the same cell and is ignored in our model. We consider only $K-$user networks where $\sqrt{K}$ is an integer, and nodes are numbered as in Figure~\ref{Cellular Network}(b). In the figure, $\sqrt{K}=6$. Since we study the performance in the asymptotic limit of the number of users, the assumption on the value of $\sqrt{K}$ is made to simplify the analysis. 

\begin{figure}[htb]
\centering
\includegraphics[scale=0.5]{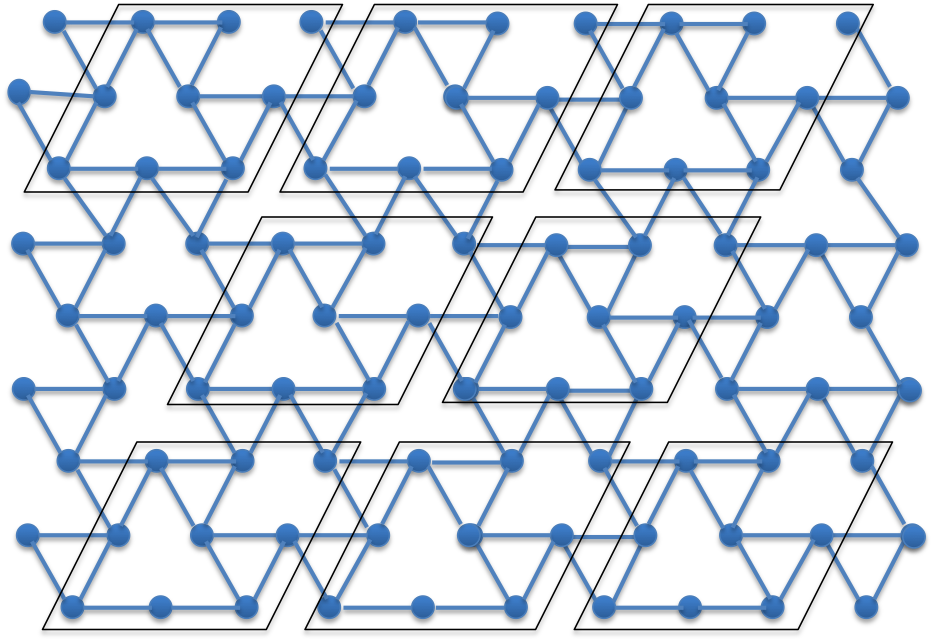} 
\protect\protect\caption{Division of cellular network into subnetworks corresponding to each macro base stations.}
\label{backhaul_division}
\end{figure}

\begin{figure}[htb]
\centering
\includegraphics[scale=0.5]{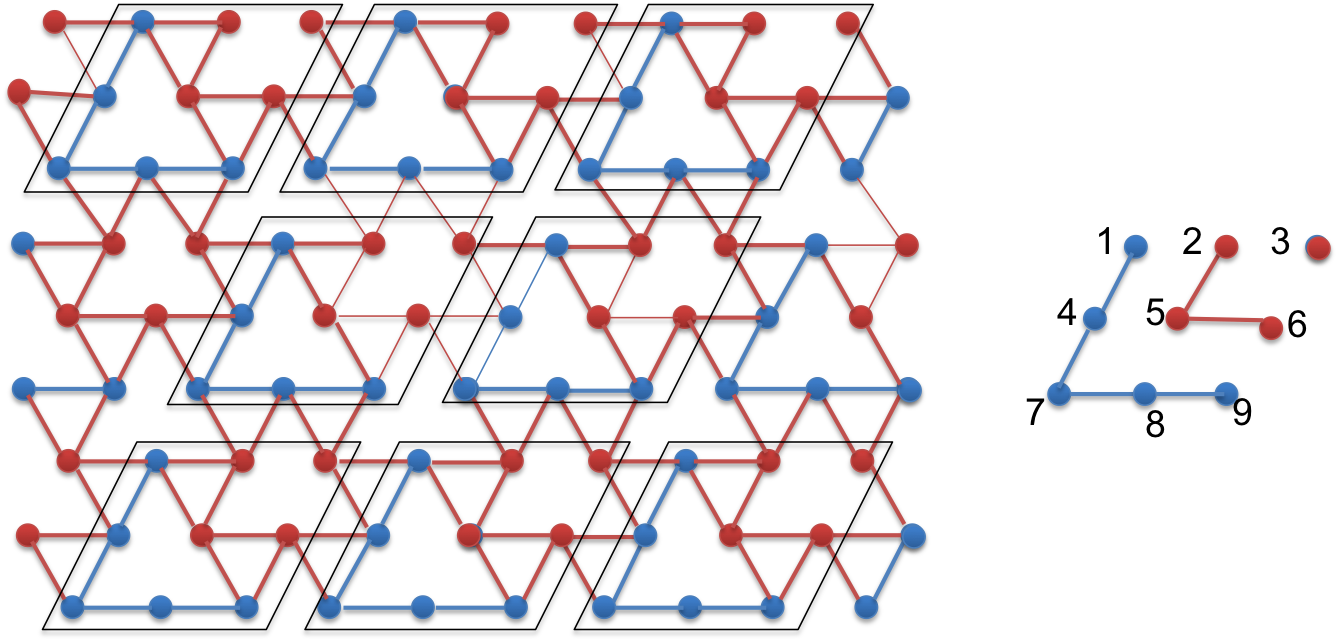} 
\protect\protect\caption{Division of cellular network into subnetworks and further division into blocks in a subnetwork.}
\label{NetworkDivision}
\end{figure}

\subsection{Backhaul layer}

Consider the division of the network into blocks of nine nodes each. Each block is associated with an MB $(i,j)$. We number the nodes in the block associated with MB $(i,j)$ as ${\mathcal S}_{(i,j)}(1)$ through ${\mathcal S}_{(i,j)}(9)$ as shown in Figure \ref{NetworkDivision}.

%We define the backhaul layer as follows. Each MB is denoted by two indices $(i,j)$. Let ${\mathcal S}_i$ denote the set of $4$ SBs where
% ${\mathcal S}_{(i,j)}=\{(2i,2j),(2i-1,2j),(2i,2j-1),(2i-1,2j-1)\}$. 
 
 Each MB $(i,j)$ is associated with a set ${\mathcal A}_{(i,j)}$ of consisting of the nine nodes it is associated with and $12$ neighboring SBs where 
 ${\mathcal A}_{(i,j)} = 
 {\mathcal S}_{(i-1,j)}(\{7,8,9\})\cup {\mathcal S}_{(i+1,j)}(\{1,2,3\})\cup {\mathcal S}_{(i,j)} \cup {\mathcal S}_{(i,j+1)}(\{1,4,7\})\cup {\mathcal S}_{(i,j-1)}(\{3,6,9\})$. Transmission from MB ${(i,j)}$ to any SB in ${\mathcal S}_{(i,j)}$ causes interference at neighboring $12$ MTs. As before, $N$ denotes the number of antennas at each MB. The channel vector between MB $(i,j)$ and SB $i'$ at time-slot $t$ is denoted by $ \mathbf{h}^{B}_{(i,j)\times i'}(t)$. The channel model for backhaul layer is given by $ \mathbf{h}^{B}_{(i,j)\times i'}(t)\neq 0$ iff  $i'\in {\mathcal A}_{(i,j)}$. 

%Let the channel gain matrix corresponding to MB $(i,j)$, $\mathbf{H}^{B}_{(i,j)}(t) \in \mathbb{C}^{(N) \times (4+8)}$ in the backhaul layer where the $j$th column corresponds to the channel coefficients from MB $(i,j)$ to SB $(i',j')$. Also let ${\mathbf x}_{i}^{B}(t) \in \mathbb{C}^{N\times 1}$ to be the transmitted signal vector from MB $i$ and $z_{k}^{B}(t)$ denotes the additive white.

\subsection{Main result}

 We show that the theoretical lower bound of half imposed by the half-duplex nature of the SBs can be achieved using zero-forcing schemes by using as few as $13$ antennas. We note that this does not require any cooperation between the MBs but does require that linear combinations be sent by the MBs to SBs to zero-force the interference at the MTs.

\begin{thm}
With $N\geq 13$, per user DoF of half can be achieved using only zero-forcing schemes. \end{thm}

\begin{IEEEproof}

First note that in each sub-block of nine nodes, i.e., the set ${\cal S}_{(i,j)}$, by deactivating SB-MT pairs corresponding to  ${\cal S}_{(i,j)}(\{2,3,5,6\})$ in an even time-slot, we are left with a non-interfering block of five nodes in each sub-block ${\cal S}_{(i,j)}(\{1,4,7,8,9\})$ and similarly by deactivating SB-MT pairs corresponding to ${\cal S}_{(i,j)}(\{1,4,7,8,9\})$ in an odd time-slot, we are left with a non-interfering block of four nodes ${\cal S}_{(i,j)}(\{2,3,5,6\})$. Thus, if these messages can be delivered by the MB $(i,j)$, we can achieve a per user DoF of $1/2$.

In the backhaul layer, at each MB, for transmitting the message to an SB in an odd time-slot, we need at least five antennas for transmitting the message to SBs ${\cal S}_{(i,j)}(\{1,4,7,8,9\})$ and eight antennas to ZF interference at ${\mathcal S}_{(i-1,j)}(\{7,8,9\})\cup {\mathcal S}_{(i+1,j)}(\{1\})\cup {\mathcal S}_{(i,j)} \cup {\mathcal S}_{(i,j+1)}(\{1,4,7\})\cup {\mathcal S}_{(i,j-1)}(\{9\})$. Similarly in each even time-slot, we need four antennas for transmitting messages and four more for ZF interference.

\end{IEEEproof}

\section{Conclusions}\label{sec:conc}

In this paper we analyzed the DoF for a linear Wyner type model for a two-tier heterogeneous wireless network. The system model, while being somewhat simplistic, does capture important features of such networks, and leads to new insights for designing more realistic cellular networks. For the linear two-tier heterogeneous wireless network, we showed that with sufficient number of antennas, a puDoF of 1/2 can be achieved using only zero-forcing schemes by sending linear combinations from MBs to their corresponding SBs so that the interference at each MT is zero-forced. These insights apply to a more realistic hexagonal cellular network where a puDoF of 1/2 can be achieved with a sufficient number of antennas using only interference avoidance schemes.

\section{Acknowledgment}
This research was supported in part by the US NSF under grant CNS 14-57168, through the University of Illinois at Urbana-Champaign, and by the Academy of Finland under grant 284801.

%We use these insights to conduct a DoF analysis for a more realistic two-dimensional cellular model considered in \cite{bande} for the transmission layer and a corresponding hexagonal cellular model for the backhaul layer. We show that a per user DoF of half can be achieved using only zero-forcing schemes even in the cellular model.

%In this paper we analyzed the DoF for a Wyner type model for a two-tier heterogeneous wireless network. The system model, while being somewhat simplistic, does capture some interesting features of the network, which could lead to new insights for designing such networks. In future work, we plan to conduct a DoF analysis for a more realistic two-dimensional cellular model considered in \cite{bande} for the transmission layer and a corresponding two-dimensional model for the backhaul layer.

\bibliographystyle{IEEEtran}

\end{document}